\documentclass[11pt,a4paper]{article}
\usepackage[utf8]{inputenc}
\usepackage[T1]{fontenc}
\usepackage{lmodern}
\usepackage{graphicx,amsmath,amssymb,booktabs,longtable,array}
\usepackage[margin=18mm]{geometry}
\usepackage{cite}
\usepackage[font=small,labelfont=bf]{caption}
\usepackage[section]{placeins}
\usepackage{xurl}
\usepackage[hidelinks]{hyperref}
\title{When a high-mountain slope failure cascades downstream:\\physical footprint and evolving exposure\\of the 2026 Gyirong mixed rock--ice cascade}
\author{Rui Li}
\date{}
\begin{document}
\maketitle
\begin{abstract}
The 26 August 2026 Gyirong cascade connected a high-mountain failure to developed downstream valleys. Event-bracketing observations delineate a preferred changed/source surface of 1.01 km$^2$, within alternative interpreted envelopes of 0.49--1.84 km$^2$, and a representative 21.8 km source-to-port route descending 3.40 km. Across two precipitation products, five antecedent windows and six fixed spatial supports, all 60 matched-year ranks remain below the 90th-percentile wet threshold. The source-nearest 7-day temperature mean of 9.43$^{\circ}$C exceeds all 25 matched years from 2001--2025. Positive temperature anomalies extend to every tested support, but ranks vary from the 88th percentile to above all historical values. Within the subsequently affected 37.354 km$^2$ UNOSAT footprint, modelled built-up surface increases from 0.036 to 0.441 km$^2$ between 1975 and 2020; its fraction of the fixed area rises from 0.096\% to 1.181\%. Delivered rapid maps contain 695 building points and 15.671 km of roads graded destroyed. These measurements connect the physical footprint with environmental context and historical exposure development. They complement existing process reconstruction while leaving the initiation mechanism and the mechanical role of warming unresolved.
\end{abstract}
\section{Introduction}
High-mountain failures can affect settlements far below their source when steep channels connect a collapsing slope to a populated valley. The resulting disaster combines a physical event with the buildings and infrastructure present along its path. Research on cascading hazards has established the importance of this connectivity, while event studies such as the 2021 Chamoli disaster show how multiple observations can constrain the movement of rock, ice and sediment.\cite{kirschbaum2019hma,mani2023cascades,shugar2021chamoli} Reconstructing the footprint, however, does not by itself identify the trigger or explain how downstream exposure developed.
These distinctions matter when satellite observations provide most of the spatial evidence. Optical images may resolve parts of a changed surface while clouds conceal its margins. Radar records surface-scattering changes under different observation conditions. Gridded meteorological products describe conditions over areas much larger than individual slope structures, and historical settlement products estimate built-up surface rather than occupancy.\cite{drusch2012sentinel2,torres2012sentinel1,munozsabater2021era5land,pesaresi2024ghsl} Combining these observations is useful only if the quantity represented by each remains clear. Otherwise, a mapped surface can be mistaken for a detached body, unusual warmth for a mechanical cause, or exposure growth for a proportional increase in losses.
On 26 August 2026, a mixed rock--ice failure in Nepal propagated toward Gyirong Port and Timure at the China--Nepal border. Guo et al.\cite{guo2026gyirong} reconstructed the connected event using satellite images, field photographs, seismic records, monitoring video, glacier surface velocities and ERA5-Land meteorology. They documented progressive entrainment, sustained glacier motion and possible short-lived activity before the main event, while leaving the immediate trigger unresolved. Their main paper and Supplemental Information also examine long-term and antecedent climate. The connected cascade and its warming background are therefore shared elements of the two studies.
Here we ask what the available observation record establishes about the event's physical footprint and downstream built-up exposure, and which interpretations of initiation remain unsupported. We combine explicit alternatives for the interpreted source surface and a representative route with matched-year hydroclimate comparisons, fixed spatial-support tests, delivered damage grades and historical built-up estimates within fixed downstream geometries. The additional measurements place environmental conditions, observed impacts and the development of exposure on separate quantitative bases. This event analysis provides evidence relevant to mountain geomorphology, hydroclimate interpretation and infrastructure planning without extending the observations into an untested rupture model.

\section{Results}
\subsection{Source-sector change and a connected downstream corridor}
\begin{figure}[!htbp]
\centering
\includegraphics[width=\textwidth,height=0.64\textheight,keepaspectratio]{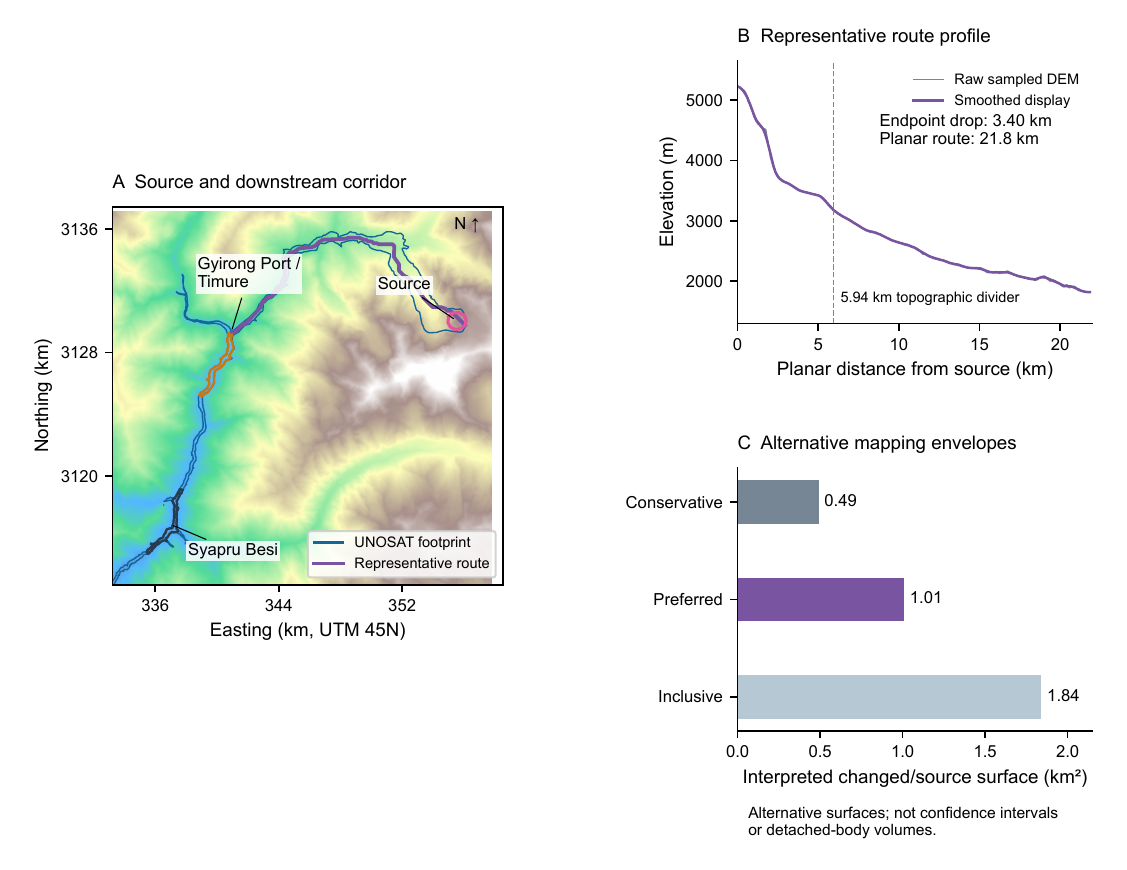}
\caption{\textbf{Physical footprint of the Gyirong cascade.} (A) Preferred source surface, representative route and downstream event-product geometries over Copernicus DEM elevation colours, focused on the source--Timure--Syapru Besi corridor. The full UNOSAT extent is shown in Figure~\ref{fig:4}. Pink marks the source, purple the route, orange the Timure zone and dark blue-grey the Syapru Besi zone; the blue outline is the UNOSAT footprint. (B) Raw DEM samples at 60~m spacing and a 21-sample quadratic Savitzky--Golay display curve. The 5.94~km divider is the first sampled elevation at or below 3200~m, not a flow-regime boundary. The planar route length differs from the approximately 22.7~km sampled profile arc. (C) Alternative interpreted changed/source surfaces; bars are not confidence intervals or detached volumes. Data: interpreted geometries, Copernicus DEM, UNOSAT and Copernicus EMSR927. Contains modified Copernicus data (2026); DEM source: Copernicus DEM GLO-30, ESA/Airbus.}
\label{fig:1}
\end{figure}

Event-bracketing observations identify a changed/source surface near 28.289$^{\circ}$ N, 85.530$^{\circ}$ E, connected to a channel descending toward Gyirong Port/Timure (Figure~\ref{fig:1}). The preferred interpreted surface covers 1.01~km$^2$; conservative and inclusive envelopes cover 0.49 and 1.84~km$^2$, respectively. These alternatives describe uncertain surface interpretation under incomplete visibility. They are neither confidence intervals nor estimates of detached-body area or volume. External spatial information was consulted before the final edges were fixed, so their agreement with previous mapping is not a blind perimeter validation.

The representative source-to-port centreline is 21.8~km long in plan and descends approximately 3.40~km between its sampled endpoints. The arc length of the sampled terrain profile is approximately 22.7~km. These lengths represent different geometric quantities, and the centreline is not the trajectory of a tracked particle. A divider at 5.94~km marks the first sampled location at or below 3200~m elevation. It is a topographic display descriptor and does not identify a transition in flow regime. Within the preferred envelope, the median sampled elevation is 4991~m and the median slope is 34.8$^{\circ}$, with predominantly west- and northwest-facing terrain (Figure~\ref{fig:s1}; Table~\ref{tab:s1}).

\begin{figure}[!htbp]
\centering
\includegraphics[width=\textwidth,height=0.64\textheight,keepaspectratio]{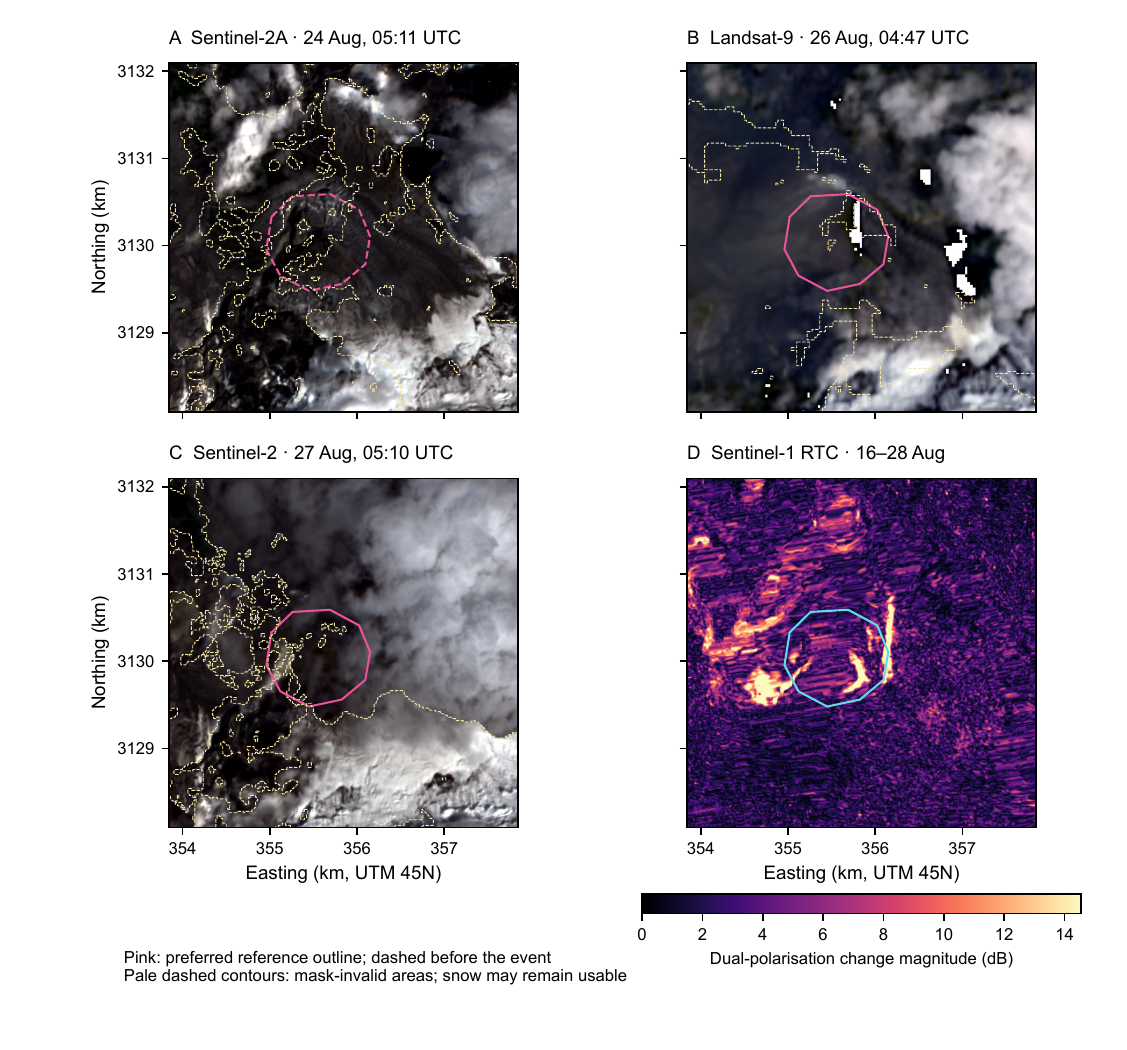}
\caption{\textbf{Cloud-limited source-sector evidence on a common map extent.} (A) Sentinel-2A S2A\_45RUM\_20260824\_1\_L2A, 24 August 2026 at 05:11:03.721 UTC. Its dashed pink outline is the preferred post-event-derived boundary shown retrospectively, not a prospective delineation. (B) Landsat-9, 26 August at 04:47:48 UTC. (C) Sentinel-2, 27 August at approximately 05:10 UTC. Solid pink outlines provide the same interpreted reference in the post-event optical views. Pale dashed contours delimit quality-mask-invalid areas; white pixels can indicate missing optical data. Snow-class pixels may conceal terrain despite passing the selection rule. (D) Sentinel-1 RTC backscatter-change magnitude between 16 and 28 August, with the same boundary in cyan. Magnitude is the nonnegative Euclidean length of the VV and VH changes in decibels, clipped at the 99th spatial percentile. It measures neither coherence nor displacement. All four axes have equal physical dimensions and the same 4~km extent in UTM 45N; the colourbar occupies separate space. The 8~m display sampling adds no native resolution. Clouds and shadow preclude reliable fine-feature callouts in the optical panels. Cloud and shadow masks indicate obscured pixels. Contains modified Copernicus Sentinel data (2026); Landsat imagery courtesy of USGS/NASA.}
\label{fig:2}
\end{figure}

The multisensor comparison shows why source delineation remains interpretive (Figure~\ref{fig:2}). Its pre-event panel is Sentinel-2A scene S2A\_45RUM\_20260824\_1\_L2A, acquired on 24 August at 05:11:03.721 UTC. Landsat-9 on 26 August and Sentinel-2 on 27 August provide post-event optical views, but cloud, shadow and missing pixels interrupt the comparison. The preferred outline on the pre-event image is a retrospective reference. Sentinel-1 radiometric terrain correction (RTC), which accounts for terrain effects on radar brightness, permits comparison of event-bracketing backscatter. The displayed dual-polarisation change magnitude supports a changed source sector; it is not a displacement or coherence measurement. The optical pixels do not justify assigning detailed fracture or material-loss labels to the obscured surface.

Six separately selected Sentinel-2 snapshots provide a limited pre-event visual record (Figure~\ref{fig:s2}; Table~\ref{tab:s2}). They reveal no unambiguous large-scale surface change in the visible portions of the selected views. The final snapshot is the distinct Sentinel-2B scene acquired on 24 August at 05:00:49.736 UTC, 45.85~h before the 26 August 02:52 UTC event reference. Approximately 62.2\% of covered preferred-source pixels meet the selection usability rule on its 20~m grid; this fraction includes snow/ice-class pixels and is not a guarantee of unobscured ground. The snapshots leave an observation gap during the final event approach and do not demonstrate absent motion or evaluate warning capability.

\subsection{Antecedent warmth extends beyond the source-nearest cell}
\begin{figure}[!htbp]
\centering
\includegraphics[width=\textwidth,height=0.64\textheight,keepaspectratio]{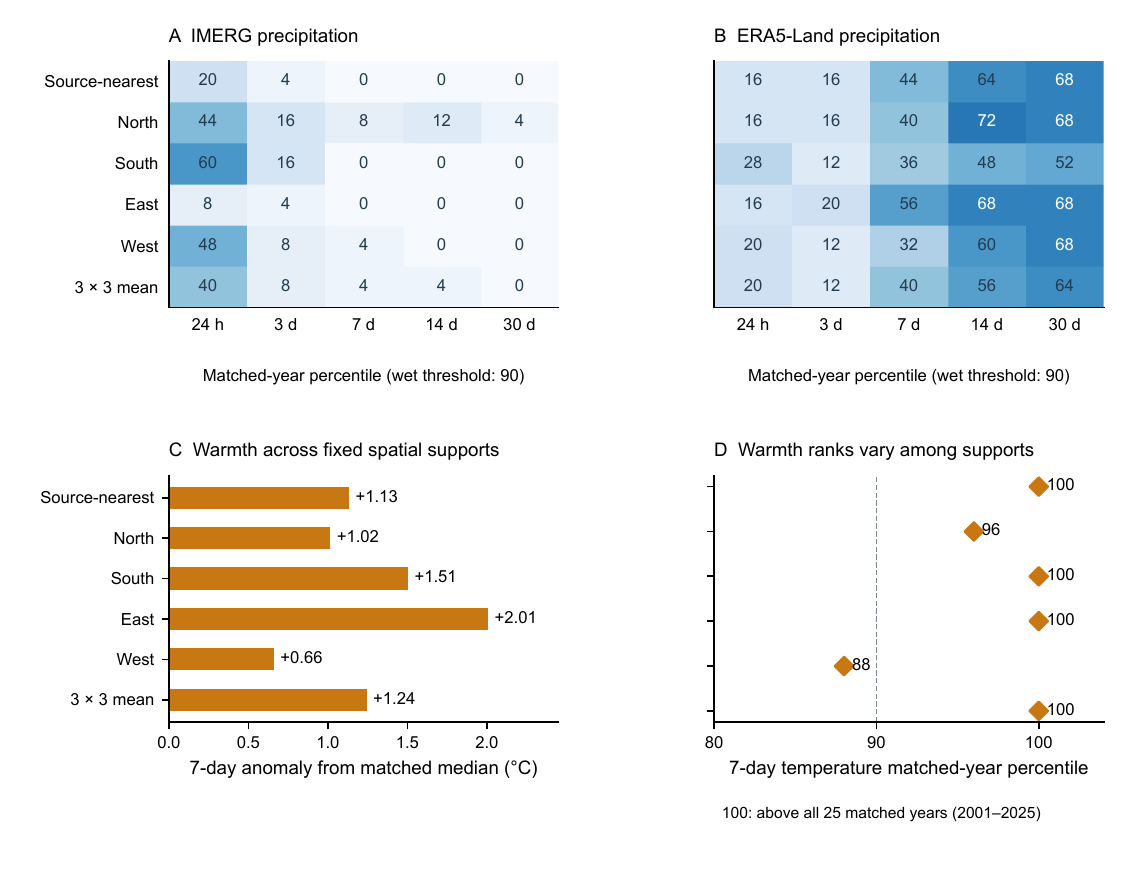}
\caption{\textbf{Antecedent hydroclimate across fixed spatial supports.} (A,B) Empirical matched-year precipitation percentiles for IMERG and ERA5-Land at the source-nearest cell, cardinal neighbours and fixed nine-cell arithmetic mean. Each support has its own 2001--2025 distribution for each window. Numbers and common colour scaling encode percentile, not rainfall amount; all 60 values are below the 90th-percentile wet threshold. (C) The 7-day mean ERA5-Land temperature anomaly relative to the median of its support-specific 25 historical means. (D) Corresponding temperature percentiles. A value of 100 means above all sampled historical means, not a return period. Only the source-nearest, south, east and nine-cell means meet that description. IMERG ends at 02:30 UTC on 26 August, ERA5-Land at 02:00 UTC. Event completeness is 100\%. Historical/event IMERG versions are V07B/V07C; the event ERA5-Land series includes preliminary ERA5-Land-T. Product-version effects are unquantified. Tables~S3 and S4 give cell coordinates, amounts and baseline values. Data: NASA GPM IMERG and Copernicus Climate Change Service ERA5-Land.}
\label{fig:3}
\end{figure}

Precipitation did not reach the declared wet-extreme threshold in either product over the tested antecedent windows (Figure~\ref{fig:3}A,B). At the source-nearest cells, IMERG totals for 24~h, 3, 7, 14 and 30~days were 4.0, 8.3, 16.8, 44.6 and 103.4~mm, compared with 5.7, 20.2, 74.5, 194.0 and 422.0~mm in ERA5-Land. The disagreement in absolute accumulation grows over longer windows. Matched against the same calendar windows in 2001--2025, the respective ranks range from below all 25 historical values to the 20th percentile for IMERG, and from the 16th to the 68th percentile for ERA5-Land (Figure~\ref{fig:s3}).

The same test across the source-nearest cell, four cardinal neighbours and a fixed nine-cell mean gives 60 product--support--window comparisons. All remain below the 90th-percentile wet threshold: ranks span 0--60 for IMERG and 12--72 for ERA5-Land (Table~\ref{tab:s3}). Here spatial support means the cell or group of cells over which a quantity is calculated. Each nine-cell mean was compared with its own historical series, rather than averaging cell percentiles. Agreement on this ranking result does not establish accurate precipitation amounts over the source slope or eliminate sub-grid rainfall variability.

The source-nearest ERA5-Land 7-day mean temperature was 9.43$^{\circ}$C, 1.13$^{\circ}$C above the median of its 25 matched historical means and about 0.15$^{\circ}$C above their maximum. Positive anomalies extend to all five additional spatial supports, ranging from 0.66 to 2.01$^{\circ}$C above each support's historical median (Figure~\ref{fig:3}C; Table~\ref{tab:s4}). Their matched ranks vary: the source-nearest, south, east and nine-cell mean exceed all 25 historical values, whereas the north and west rank at the 96th and 88th percentiles (Figure~\ref{fig:3}D). Thus, antecedent warmth extends beyond a single selected grid cell, but an above-all-years statement does not hold everywhere.

Preliminary ERA5-Land-T values and the historical/event IMERG V07B/V07C difference remain possible influences on the event comparisons; their effects were not separately quantified. The temperature rankings do not establish the mechanical role of warming in this rupture.

\subsection{Historical built-up exposure within the mapped impact area}
\begin{figure}[!htbp]
\centering
\includegraphics[width=\textwidth,height=0.64\textheight,keepaspectratio]{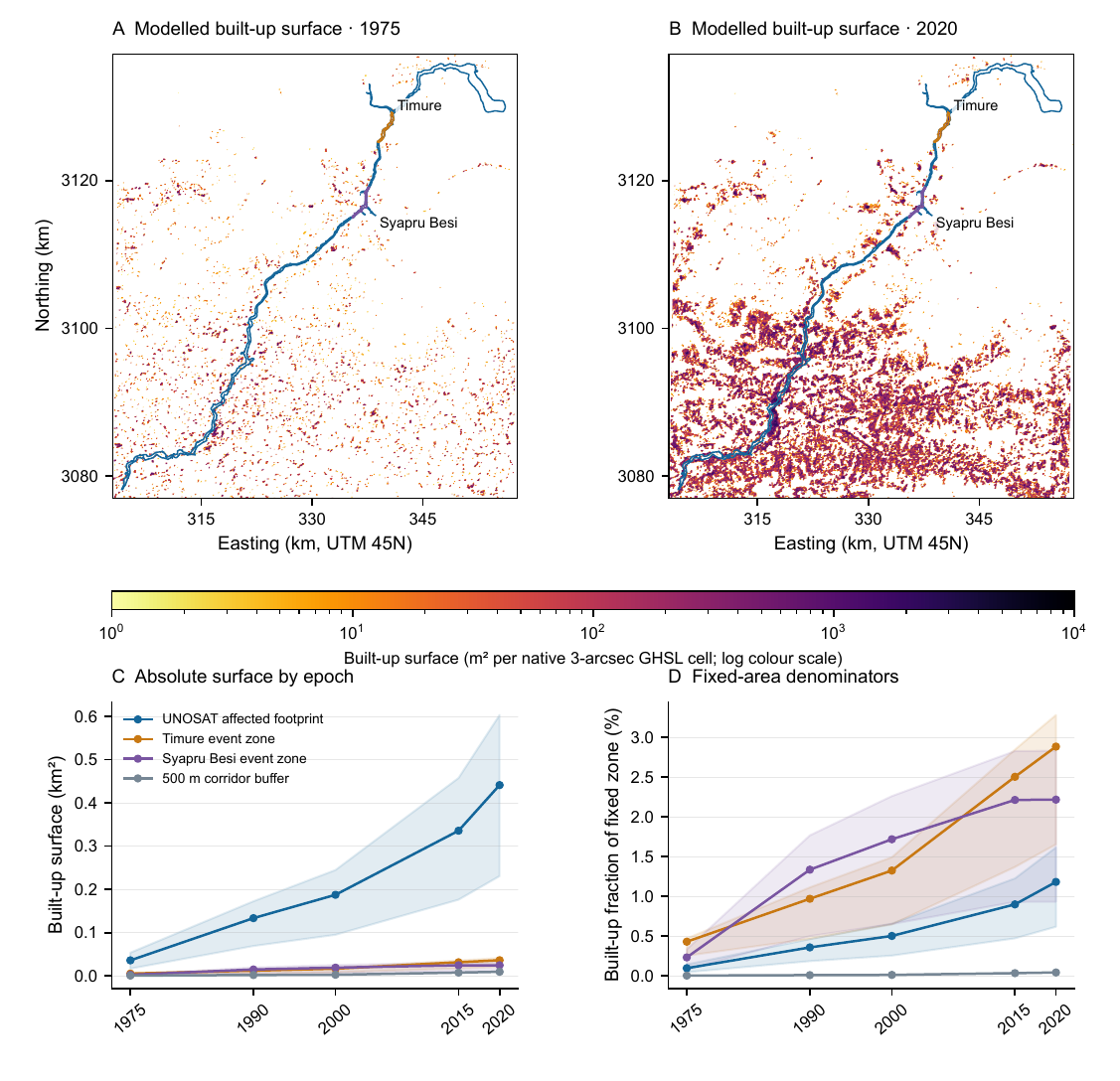}
\caption{\textbf{Historical built-up surface within fixed downstream zones.} (A,B) GHSL GHS-BUILT-S R2023A for 1975 and 2020 over the same UTM 45N extent. Colours show square metres of built-up surface per native 3-arcsec cell on an identical logarithmic scale; zero cells are uncoloured. Blue outlines show the UNOSAT affected footprint, orange the Timure EMS event polygon and purple the Syapru Besi EMS event polygon. (C) Preferred cell-centre sums for all five epochs in four fixed zones, including the 500~m route buffer. (D) The same sums divided by each full fixed polygon area, expressed as percentages. Bands span sums from a 65~m eroded-core cell-centre selection to all cells touched by the original polygon; denominators remain the original full polygon areas. The bands exclude classification and historical-model uncertainty. Zones overlap and EMS zones are not entire municipalities. These modelled historical surfaces describe exposure in the subsequently affected footprint, not individual construction dates or historical losses. Zone definitions, areas and values are in Tables~S5 and S6. Data: European Commission Joint Research Centre GHSL, UNOSAT and Copernicus EMSR927; contains modified Copernicus service information (2026).}
\label{fig:4}
\end{figure}

Historical settlement estimates show increasing built-up surface within geometries later intersected by the event (Figure~\ref{fig:4}; Tables~S5 and S6). The Global Human Settlement Layer (GHSL) represents modelled built-up surface, expressed as square metres within each raster cell, at historical epochs. Within the fixed 37.354~km$^2$ UNOSAT affected footprint, preferred sums rise from 0.03595~km$^2$ in 1975 to 0.44124~km$^2$ in 2020, an absolute increase of 0.40530~km$^2$. Relative to the same full polygon area, the estimated built-up fraction rises from 0.096\% to 1.181\%. The 1975 and 2020 maps use identical extents and native-cell units, making the spatial distribution of the estimates visible alongside their totals.

The smaller event-product zones provide complementary views of this change. In the 1.259~km$^2$ Timure/Gyirong Port zone, modelled built-up surface increases from 0.00541 to 0.03629~km$^2$, and the fraction from 0.430\% to 2.883\%. In the 1.111~km$^2$ Syapru Besi zone, corresponding estimates increase from 0.00258 to 0.02462~km$^2$, or 0.232\% to 2.216\%. A 500~m buffer around the representative route provides a fourth fixed spatial support. These overlapping zones are not independent replicates, and the emergency-map polygons do not delimit entire municipalities.

Alternative boundary sampling changes the absolute GHSL sums. Within the UNOSAT footprint, a 65~m eroded core and an all-touched-cell selection give 1975 estimates of 0.01679--0.05372~km$^2$ and 2020 estimates of 0.23067--0.60332~km$^2$. For each of the four zones, the lower 2020 estimate exceeds the upper 1975 estimate. This comparison preserves the direction of historical growth under the tested boundary choices, while excluding uncertainty in the classification and historical modelling of built-up surface. It does not reconstruct the construction date of any particular structure or establish how losses would have differed in 1975.

\begin{figure}[!htbp]
\centering
\includegraphics[width=\textwidth,height=0.64\textheight,keepaspectratio]{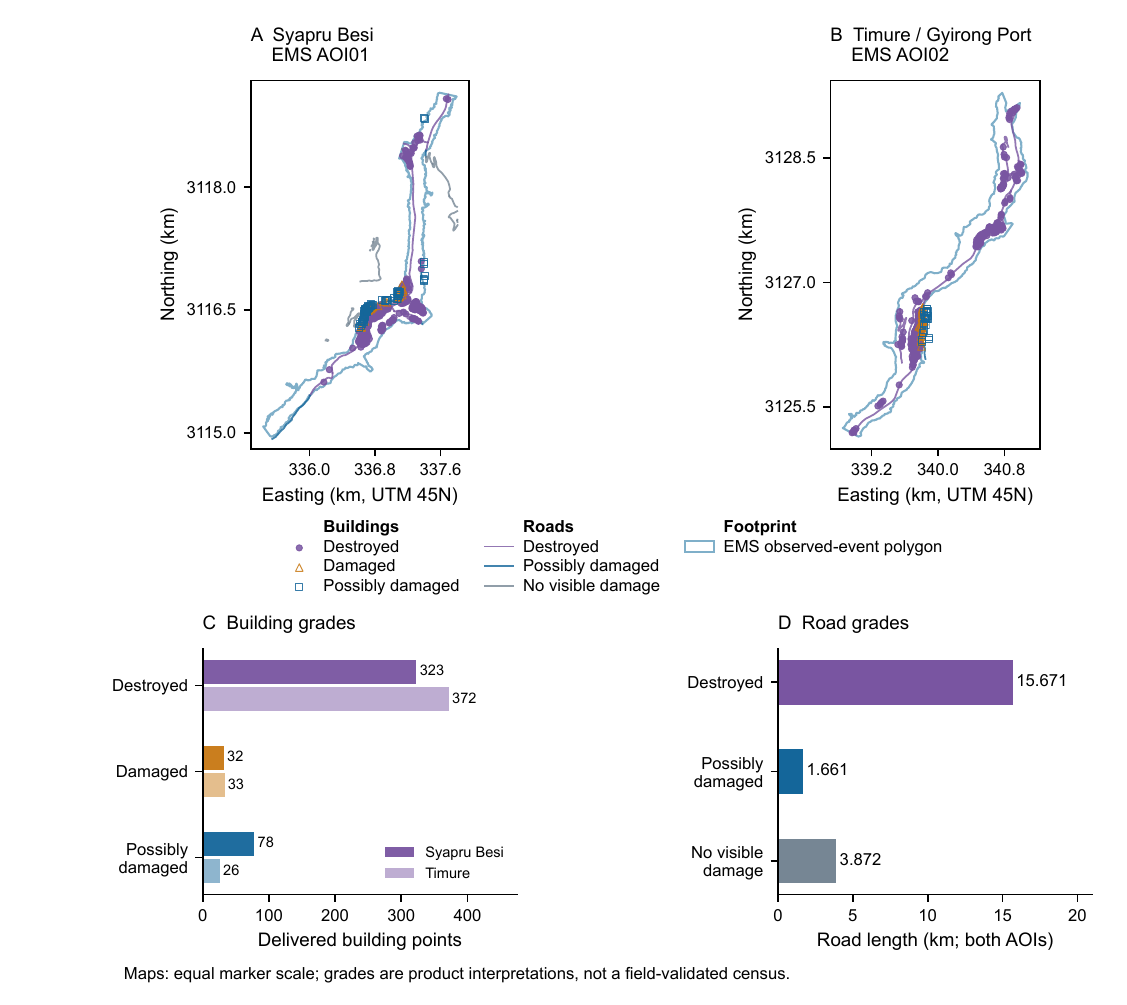}
\caption{\textbf{Delivered damage grades in two downstream mapping areas.} (A,B) EMSR927 AOI01 Syapru Besi and AOI02 Timure/Gyirong Port. Building grades use purple filled circles for destroyed, orange hollow triangles for damaged and blue hollow squares for possibly damaged, with comparable marker areas. The shared legend distinguishes building markers from road lines. Roads use the corresponding grade colours; grey denotes No visible damage. Pale-blue outlines show the EMS observed-event polygon in each panel. Coordinates are UTM 45N. (C) Delivered building-record counts by grade and area; darker bars denote Syapru Besi and lighter bars Timure. All 864 building records have coordinates. Two Syapru Besi road records graded no visible damage lack geometry and contribute no measurable length. (D) Road lengths by grade across both areas. The totals are 695 building points and 15.671~km of roads graded destroyed. These are interpreted product grades without field validation or a complete-census denominator in this study. Table~\ref{tab:s7} provides the underlying counts and geometry coverage. Data: Copernicus Emergency Management Service EMSR927 (2026); contains modified Copernicus service information (2026).}
\label{fig:5}
\end{figure}

Copernicus Emergency Management Service (EMS) grading products map substantial damage in the two downstream areas (Figure~\ref{fig:5}; Table~\ref{tab:s7}). The delivered layers contain 864 building points: 695 graded destroyed, 65 damaged and 104 possibly damaged. Destroyed points comprise 323 in Syapru Besi and 372 in Timure. Mapped roads include 15.671~km graded destroyed, with a further 1.661~km possibly damaged and 3.872~km showing no visible damage. Six transport points and two facility polygons are also graded destroyed. These are the provider's interpreted grades for delivered features, not a field-validated census of all affected assets. Historical GHSL surface and 2026 damage points have different observation units and dates; they are not joined building by building.

\section{Discussion}
The observations establish a large interpreted source-sector change, a long descending corridor and mapped downstream damage, while documenting the prior development of built-up exposure within fixed affected-area geometries. Keeping those quantities separate makes the event informative beyond source mechanics. The physical connection explains why a remote high-mountain failure can intersect developed valley locations. The historical estimates show that the built environment occupying the later footprint changed substantially before the event. Neither result requires treating settlement growth as a cause of the initial collapse.

The specific additions complement Guo et al.'s process reconstruction.\cite{guo2026gyirong} Their approximately 0.72~km$^2$ detached ice--rock body and the 1.01~km$^2$ preferred changed/source surface here are differently defined objects; their numerical difference is not evidence that either mapped area validates or refutes the other. Their approximately 22~km propagation distance and 3400~m relief agree in scale with the representative route, but shared imagery and consulted external context preclude claiming independent final-edge confirmation. Their glacier-velocity series and possible seismic activity address processes that the six selected optical snapshots cannot resolve. The present measurements add explicit surface alternatives, a two-product, multi-window hydroclimate comparison across fixed spatial supports, category-specific delivered damage totals and historical built-up estimates with spatial denominators and boundary sensitivity.

The hydroclimate results narrow an environmental description without determining a trigger. The lack of a wet-extreme rank across both products and all tested supports makes a spatially widespread antecedent precipitation extreme unsupported by these data. It does not rule out water involvement: infiltration, meltwater availability, drainage changes and local rainfall can differ from coarse-grid precipitation totals. Likewise, a warm 7-day mean is not a measurement of bed temperature, melt volume, fracture-water pressure or strength. Temperature-index positive degree days, the sum of positive daily mean temperatures, provide additional thermal context in the Supplemental Information but do not quantify source melt.\cite{wake2015pdd} Establishing a mechanical temperature pathway would require observations of the relevant material and hydrological state.

The fixed 255-cell ERA5-Land July--August series shows warming of approximately 0.121$^{\circ}$C per decade over 1951--2025, with a positive Berkeley Earth trend over its larger comparison region (Figure~\ref{fig:s4}; Table~\ref{tab:s8}). This distinction also applies to anthropogenic interpretation. Regional warming and its attribution in broader literature concern scales and questions different from those of an individual rupture.\cite{gruber2017hkh,jiang2023tibetan} The small four-model forcing comparison retained in the Supplemental Information is contextual and cannot estimate an attributable probability for this event. Direct local disturbance at the source remains insufficiently assessed because no validated historical disturbance survey exists for that geometry. The evidence assembled here does not support attributing the rupture to direct human disturbance. The mechanical contribution of regional warming remains unresolved, whereas historical growth of downstream built-up exposure is documented.

The exposure result has its own limits. GHSL epochs are modelled historical estimates, not inventories of structures observed at exact construction dates.\cite{pesaresi2024ghsl} Fixed affected-area polygons answer how much modelled built-up surface occupied the subsequently mapped footprint; they do not compare growth with unaffected places. The study therefore cannot determine whether growth was disproportionate, estimate historical disaster losses, or infer changes in hazard frequency, magnitude or source stability. Population, occupancy, vulnerability and monetary value are also outside the measured quantities. For planning, the useful distinction is between documenting development in a connected downstream corridor and assigning a quantified change in risk, which would require additional evidence.

Observation limits should also guide the use of this reconstruction. Sparse cloud-limited snapshots cannot resolve the final hours of source evolution, and the backscatter magnitude is not a source-motion test. No source volume, entrained volume or initial blockage mechanism is established here. The 533~m median of selected disturbed-footprint transects is retained with its censoring explanation in the Supplemental Information, rather than being presented as a hydraulic width. Damage totals inherit the scope and limitations of the delivered rapid maps. Together, these boundaries define what the case can support: a spatially explicit account of the physical footprint, environmental context and changing downstream exposure, with the initiation process still open.

\section{Materials and methods}
\subsection{Event reference, source interpretation and geometry}
The analysis uses existing event-bracketing Sentinel-2, Landsat-9 and Sentinel-1 observations, Copernicus DEM GLO-30, fixed interpreted geometries and delivered rapid-mapping products.\cite{drusch2012sentinel2,torres2012sentinel1,landsatCollection2,copernicusDem,unosat2026gyirong,cems2026gyirong} The reference time is 26 August 2026 at 02:52 UTC, used to align image gaps and antecedent windows, not as a directly observed moment of initial rupture. Planet disaster imagery and external event information informed interpretation; Planet pixels are not reproduced in the figures. Source envelopes represent conservative, preferred and inclusive interpretations of the changed surface.

Area and route length use WGS84/UTM zone 45N (EPSG:32645). GLO-30 was bilinearly sampled to the existing 10~m analysis grid for source terrain statistics and route profiling; this sampling does not increase native terrain information. The minimum rotated rectangle's long side and a 101$\times$101-cell square-window elevation range are reported only with their definitions (Table~\ref{tab:s1}). Source imagery in Figure~\ref{fig:2} is displayed over a common 4~km square on an 8~m display grid, with equal physical map-axis dimensions and separate colourbar space. Interpolation for display adds no spatial resolution. Optical channel stretches use the first and 99th percentiles; cloud and quality masks are shown without cloud removal. RTC change is the Euclidean magnitude of post-minus-pre VV and VH backscatter differences in decibels. It contains no phase information.

The six pre-event Sentinel-2 scenes were selected within fixed target-date windows by highest preferred-source scene-classification usability and then proximity to the target date. Classes 4, 5, 6, 7 and 11 were retained on a 20~m selection grid. Coverage, usable fraction and scene-wide cloud fraction have different denominators; the displayed sequence uses 10~m sampling. Exact scene identities and times are provided in Table~\ref{tab:s2}. The source boundary is retrospective in every pre-event panel.

\subsection{Matched antecedent conditions and spatial supports}
IMERG Late precipitation is the sum of rates multiplied by interval durations, ending at 02:30 UTC on 26 August.\cite{huffman2020imerg,imergLateV07} ERA5-Land precipitation and hourly mean 2~m temperature end at 02:00 UTC.\cite{munozsabater2021era5land,era5landDataset} Windows of 24~h, 3, 7, 14 and 30~days are compared with their calendar-matched counterparts in each year from 2001 to 2025. The empirical percentile is $100(n_{<}+0.5n_{=})/25$, where the counts refer to historical values below or equal to the event value. Ranks at 0 or 100 mean below or above all sampled years, not population extremes or return periods. Overlapping windows and neighbouring supports are dependent.

The fixed source-nearest, north, south, east and west supports each contain one 0.1$^{\circ}$ cell; the sixth is the arithmetic mean of the fixed 3$\times$3 block. Product grids differ in cell-centre position (Tables~S3 and S4). Historical comparisons are recalculated separately from the saved 25-value distribution for every support. Temperature anomalies subtract the support-specific median of matched 7-day means. All retained event comparisons have complete temporal coverage. Historical/event IMERG files are V07B/V07C, and the event ERA5-Land record includes preliminary ERA5-Land-T, expver 0005. No test here isolates the effects of these version differences. Temperatures and positive degree days are not lapse-rate corrected. Long-term regional supports, regression assumptions and the limited CMIP6/DAMIP comparison are specified in the Supplemental Information.\cite{gillett2016damip,rohde2020berkeley}

\subsection{Historical built-up surface and delivered damage}
GHSL GHS-BUILT-S R2023A values at 3-arcsec sampling were summed for 1975, 1990, 2000, 2015 and 2020 within four fixed geometries: the union of the delivered UNOSAT affected footprint, the EMS Timure and Syapru Besi observed-event polygons, and a 500~m buffer around the representative centreline.\cite{pesaresi2024ghsl,unosat2026gyirong,cems2026gyirong} Preferred sampling selects cells by their centres. Alternative sums use cell centres inside a 65~m eroded polygon and all cells touched by the original polygon. The original native raster values already represent square metres of built-up surface and are not multiplied by pixel area. Fractions divide these sums by the full fixed polygon area in EPSG:32645, including for the sensitivity estimates; they do not use selected-cell area as a denominator. These tests address boundary sampling, not product classification error.

Building records and road geometries from EMSR927 AOI01 and AOI02 were grouped by the delivered damage grades. All 864 building records have map coordinates. Road lengths were measured from available geometries in EPSG:32645; two AOI01 records graded no visible damage lack geometry and contribute no measurable length. The plotting tables distinguish delivered records from measurable geometries. No field validation, population conversion or monetary-loss model was applied. Historical exposure and event damage are compared at the level of their fixed spatial supports, without asserting individual feature correspondence.

\section{Data and code availability}
Upstream observations and products are available from the Copernicus Data Space Ecosystem (Sentinel), USGS/NASA (Landsat), the Copernicus Climate Change Service (ERA5-Land; \url{https://doi.org/10.24381/cds.e2161bac}), NASA GES DISC (IMERG Late V07; \url{https://doi.org/10.5067/GPM/IMERG/3B-HH-L/07}), the European Commission Joint Research Centre (GHSL), UNOSAT, Copernicus EMSR927, Berkeley Earth and the Pangeo CMIP6 catalogue. Dataset references and the Supplemental Information specify the products and measurement supports. Planet disaster imagery is available through its provider catalogue and remains subject to its separate reuse terms. Processed data and analysis code have not been deposited in a public archive.

\section*{Data acknowledgments}
Terrain displays use Copernicus WorldDEM-30 \textcopyright{} DLR e.V. 2010--2014 and \textcopyright{} Airbus Defence and Space GmbH 2014--2018 provided under COPERNICUS by the European Union and ESA; all rights reserved. The organisations in charge of the Copernicus programme by law or by delegation do not incur any liability for any use of the Copernicus WorldDEM-30. Sentinel, Landsat, GHSL and rapid-mapping product credits are given in the figure captions.

\clearpage
\label{main_end}
\appendix
\setcounter{figure}{0}
\setcounter{table}{0}
\renewcommand{\thetable}{S\arabic{table}}
\renewcommand{\theHtable}{supp.\arabic{table}}
\renewcommand{\thefigure}{S\arabic{figure}}
\renewcommand{\theHfigure}{supp.\arabic{figure}}
\section*{Supplemental Information}
\phantomsection\label{supp_start}
\section*{Supplemental Materials and Methods}
\subsection*{Source geometry and observation support}
Source-envelope coordinates and source calculations use the conservative, preferred and inclusive geometries. Their exact projected areas are 0.4906625, 1.0093375 and 1.84125~km$^2$. Terrain was bilinearly resampled from GLO-30 to the existing 10~m grid; the preferred-source median elevation is 4991.1973~m, slope 34.7675$^{\circ}$ and circular aspect 280.7786$^{\circ}$. West and northwest aspect classes contain 48.2384\% and 33.7511\% of preferred-envelope cells. The 1151.3558~m rectangle measure is the longest side of the minimum rotated bounding rectangle, not a maximum pairwise separation. Median local relief is 839.7549~m using a 101$\times$101-cell square window. The boundaries were not wholly blind to external spatial information and do not represent a detached volume.\cite{copernicusDem,guo2026gyirong}

Figure~\ref{fig:2}'s Sentinel-2A acquisition and Figure~\ref{fig:s2}'s final Sentinel-2B acquisition differ by 10~min 13.985~s. The 45.852851~h gap uses the latter acquisition and the 26 August 02:52:00 UTC reference. The final S2B selected-source fraction is 62.1908\% on the 20~m selection grid. These values are not scene-wide cloud percentages or guarantees of visible terrain. Retained scene classes are vegetation, non-vegetated surface, water, unclassified and snow/ice (classes 4, 5, 6, 7 and 11). Cloud/shadow masking and optical stretches do not restore obscured information. PlanetScope observations at 26 August 05:44:56 UTC and later dates informed interpretation but Planet pixels are not reproduced here.

Sentinel-1 comparison uses the 16 and 28 August ascending RTC pair. For positive finite VV and VH intensities, each difference is post-event minus pre-event $10\log_{10}$ backscatter, and the magnitude is $\sqrt{(\Delta VV_{\rm dB})^2+(\Delta VH_{\rm dB})^2}$. It has no phase, coherence or displacement interpretation. Figure~\ref{fig:2} uses a 4~km extent with identical 8~m display sampling across panels. The separate six-scene sequence uses a smaller extent at 10~m sampling.

\subsection*{Route, transects and limits of source-motion evidence}
The representative centreline length is 21.84342488~km in EPSG:32645. Raw DEM sampling gives a 22.70963939~km arc length; the saved rounded samples sum to 22.70981444~km. Both round to 22.710~km. Raw endpoint elevations are 5219.7251 and 1823.4246~m, giving a 3396.3005~m drop and an 8.8378$^{\circ}$ route-gradient angle. The angle uses route length, not straight-line endpoint separation. The first sampled location at or below 3200~m is at 5.94~km, with elevation 3181.2~m. Local uphill intervals remain in the sampled profile; their causes are not established.

Twelve fixed transects measure selected components of the mapped disturbed footprint (Table~\ref{tab:s9}). T01--T04 each have one endpoint inside it, making their measured component lengths censored lower bounds; T05--T12 cross both component boundaries. All four lower bounds exceed the two complete component widths occupying ranks six and seven in the full set, T07 at 388.902453~m and T06 at 677.312079~m. Their mean is 533.107266~m, so the median remains identifiable despite censoring. No finite overall maximum is estimated. These are mapped footprint-component widths along fixed directions, not water, flow or hydraulic widths. No water-width result is reported.

Feature tracking was limited to calibration with synthetic shifts, stable terrain and non-event controls. No source displacement field, source spatial-coherence statistic or acceleration test is reported. No validated historical disturbance or cryosphere-change survey exists for the source geometry. Guo et al.'s 211-scene autoRIFT/TICOI glacier-motion analysis, field imagery and possible seismic activity are external evidence and are not reclassified as results of this study.\cite{guo2026gyirong}

\subsection*{Antecedent hydroclimate and historical comparisons}
Tables~S3 and S4 expose all fixed support comparisons. Single-cell centres differ between ERA5-Land and IMERG; the nine-cell mean is the arithmetic mean of the fixed block's values before constructing the matched-year distribution. Historical ranks are never averaged. Every event comparison is temporally complete, with 25 complete matched years over 2001--2025. Precipitation windows end on 26 August at 02:30 UTC for IMERG and 02:00 UTC for ERA5-Land; hourly mean temperature uses the latter cutoff. Precipitation rate is integrated over each interval. Ranks use $100(n_<+0.5n_=)/25$, and the declared wet threshold is the 90th percentile. Product-version differences remain untested: historical/event IMERG Late is V07B/V07C, and event ERA5-Land includes preliminary ERA5-Land-T (expver 0005).\cite{huffman2020imerg,imergLateV07,munozsabater2021era5land}

Temperature anomalies in the main text, Figure~\ref{fig:3} and Table~\ref{tab:s4} subtract the support-specific median of the 25 matched 7-day means. The machine-readable table also includes the historical mean and the anomaly from that mean, explicitly named as a different descriptive statistic. The source-nearest 7-day mean exceeds its historical maximum by approximately 0.15045$^{\circ}$C. The positive-degree-day index sums positive daily mean temperatures over complete UTC days through 25 August, and is not lapse-rate corrected. Source-nearest 7, 14 and 30~day and 1 July--25 August totals are 65.94, 131.11, 279.44 and 526.02$^{\circ}$C~d, with ranks 100, 96, 88 and 96. It is a temperature index, not a source-melt estimate.\cite{wake2015pdd}

Under the linear daily 90th-percentile rule, 21--25 August forms a five-day exceedance sequence. On 22 August, the daily value of 9.106113688$^{\circ}$C exceeds the threshold of 9.105694580$^{\circ}$C by only 0.000419108$^{\circ}$C. This near-threshold value makes exact duration sensitive to threshold convention and reporting precision. The effect of preliminary product values on this duration has not been quantified. The exact duration is therefore not a main result. The parent-ERA5 terrain proxy is not a measured native ERA5-Land source-elevation mismatch and is not used for an elevation correction.

\subsection*{Regional temperature and model supports}
ERA5-Land July--August regional means use an arithmetic mean of 255 grid centres spanning 27.6--29.2$^{\circ}$ N and 84.8--86.2$^{\circ}$ E. Only July--August hourly data underpin this series; annual-mean and full-JJA trends are not inferred. Berkeley Earth uses nine 1$^{\circ}$ cells, with centres at 27.5, 28.5 and 29.5$^{\circ}$ N and 84.5, 85.5 and 86.5$^{\circ}$ E, selected within the larger 27--30$^{\circ}$ N, 84--87$^{\circ}$ E box and weighted by cosine latitude. Its input has file-history date 9 January 2025 and covers the comparison through 2024. Positive trends across the differently supported products are contextual agreement, not numerical equivalence.\cite{rohde2020berkeley}

Ordinary least-squares (OLS) trends and 95\% intervals assume independent residuals (Table~\ref{tab:s8}); the intervals omit serial-dependence corrections and full product uncertainty. CMIP6/DAMIP uses four complete model triplets selected by archive availability: one member per model, historical ALL through 2014 continued with SSP2-4.5 through 2020, paired with historical-natural forcing. Monthly means are weighted by days per month, and spatial means by cosine latitude. BCC-CSM2-MR, CNRM-CM6-1, CanESM5 and GFDL-CM4 contribute 9, 4, 1 and 9 centres, respectively. The single CanESM5 centre at 29.30136$^{\circ}$ N, 84.375$^{\circ}$ E cannot resolve mountain-scale regional variability. Exact model cells and archive identifiers are retained in the plotting data.\cite{gillett2016damip}

For 1951--2020, all four ALL trends are positive and two lie within the observed ERA5-Land OLS interval; no NAT trend is positive or inside that interval. For 1971--2020, all ALL trends are positive but none lies inside the observed interval; three NAT trends are positive and none lies inside it. The four-model spread is not a probability distribution for natural variability, and the observational/model boxes differ. This is neither formal regional detection--attribution nor event attribution.

\subsection*{Exposure and damage measurement}
The four GHSL zones are fixed raw-layer unions and the 500~m representative-route buffer (Table~\ref{tab:s5}). Their full areas are measured in EPSG:32645 and remain the denominator for preferred and sensitivity fractions. Preferred native-raster cell-centre sums were reproduced for all 20 zone--epoch combinations, along with eroded-core and all-touched alternatives. A 65~m erosion defines the core. The UNOSAT union area is 37.35419474~km$^2$ and is the fixed denominator for the historical GHSL analysis. Values are square metres per native GHSL cell, with no second pixel-area multiplication.\cite{pesaresi2024ghsl,unosat2026gyirong}

Tables~S5 and S6 report fixed areas, epoch sums, fractions, absolute changes, relative changes and sampling bands. The bands exclude classification error. The zones overlap, and no unexposed control area was analysed. Built-up surface is not population, occupancy, vulnerability, monetary value or disaster loss. All building records have coordinates. Two Syapru Besi road records graded no visible damage lack geometry and contribute no measurable length; only valid geometries appear on maps. No damage grade was inferred from an OpenStreetMap feature.\cite{cems2026gyirong}

\begin{figure}[!htbp]
\centering
\includegraphics[width=\textwidth,height=0.64\textheight,keepaspectratio]{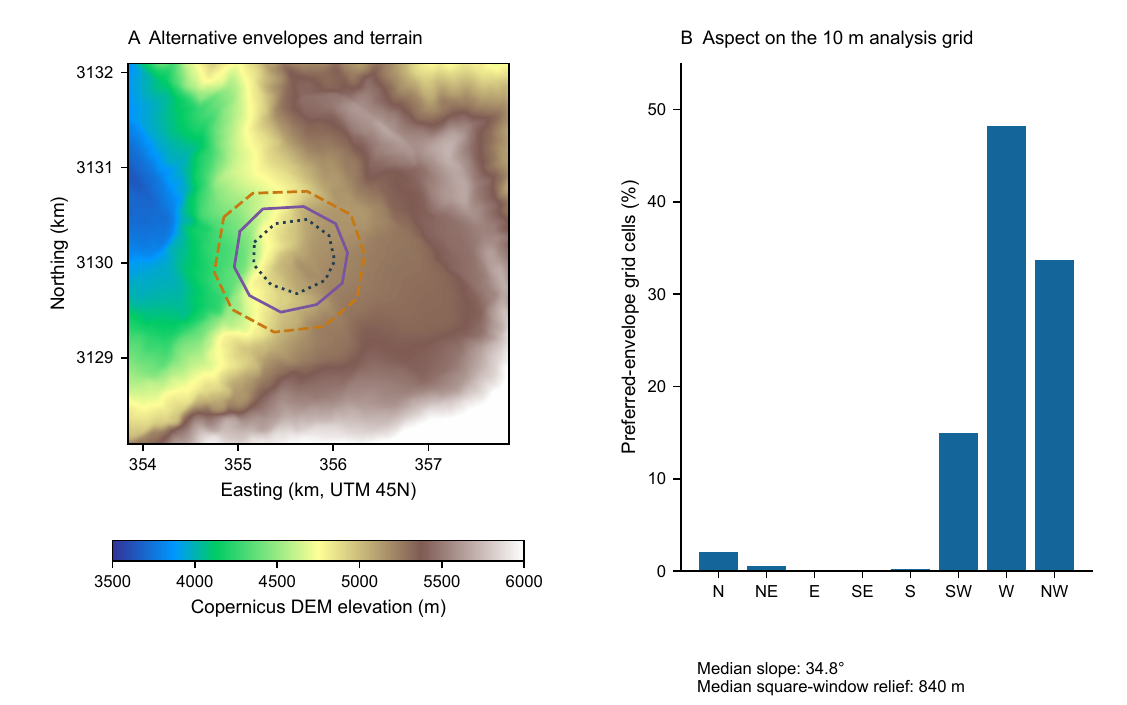}
\caption{\textbf{Source envelopes and terrain context.} (A) Conservative dotted, preferred solid and inclusive dashed interpreted surface outlines over Copernicus DEM. The displayed DEM is resampled; source statistics use the original 10~m analysis grid derived from GLO-30. (B) Aspect-class proportions within the preferred envelope. Median slope is 34.8$^{\circ}$ and median 101$\times$101-cell square-window relief is 840~m. The source alternatives express surface-interpretation choices, not uncertainty in detached volume. Table~\ref{tab:s1} gives full metrics. Data: interpreted envelopes and Copernicus DEM GLO-30, ESA/Airbus.}
\label{fig:s1}
\end{figure}

\begin{figure}[!htbp]
\centering
\includegraphics[width=\textwidth,height=0.64\textheight,keepaspectratio]{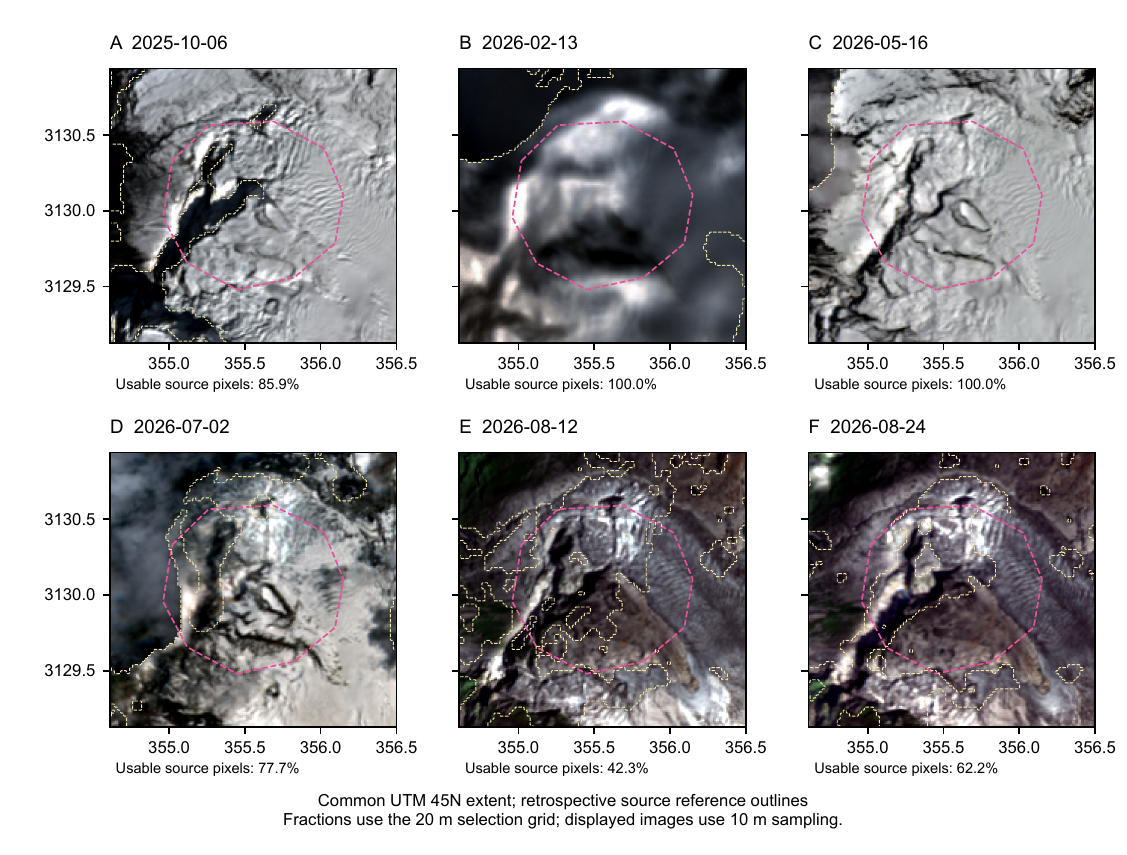}
\caption{\textbf{Six selected pre-event source views.} (A--F) Sentinel-2 scenes selected within fixed target-date windows. Dates are actual acquisitions, not the nominal target offsets. All panels use the same map extent and 10~m display sampling; dashed pink source outlines are retrospective references. Pale dashed contours mark quality-mask-invalid areas. The printed usable-source fractions use covered preferred-source pixels on the 20~m selection grid and retain scene-classification classes 4, 5, 6, 7 and 11, including snow/ice. Easting and northing axes are kilometres in UTM 45N. The last scene is S2B\_45RUM\_20260824\_0\_L2A at 05:00:49.736 UTC, distinct from Figure~\ref{fig:2}A's S2A scene; its gap to the event reference is 45.85~h. Visible pixels do not show unambiguous large-scale change, but cloud, shadow, seasonal snow and temporal gaps leave other changes unresolved. Table~\ref{tab:s2} lists exact identities. Contains modified Copernicus Sentinel data (2025--2026).}
\label{fig:s2}
\end{figure}

\begin{figure}[!htbp]
\centering
\includegraphics[width=\textwidth,height=0.64\textheight,keepaspectratio]{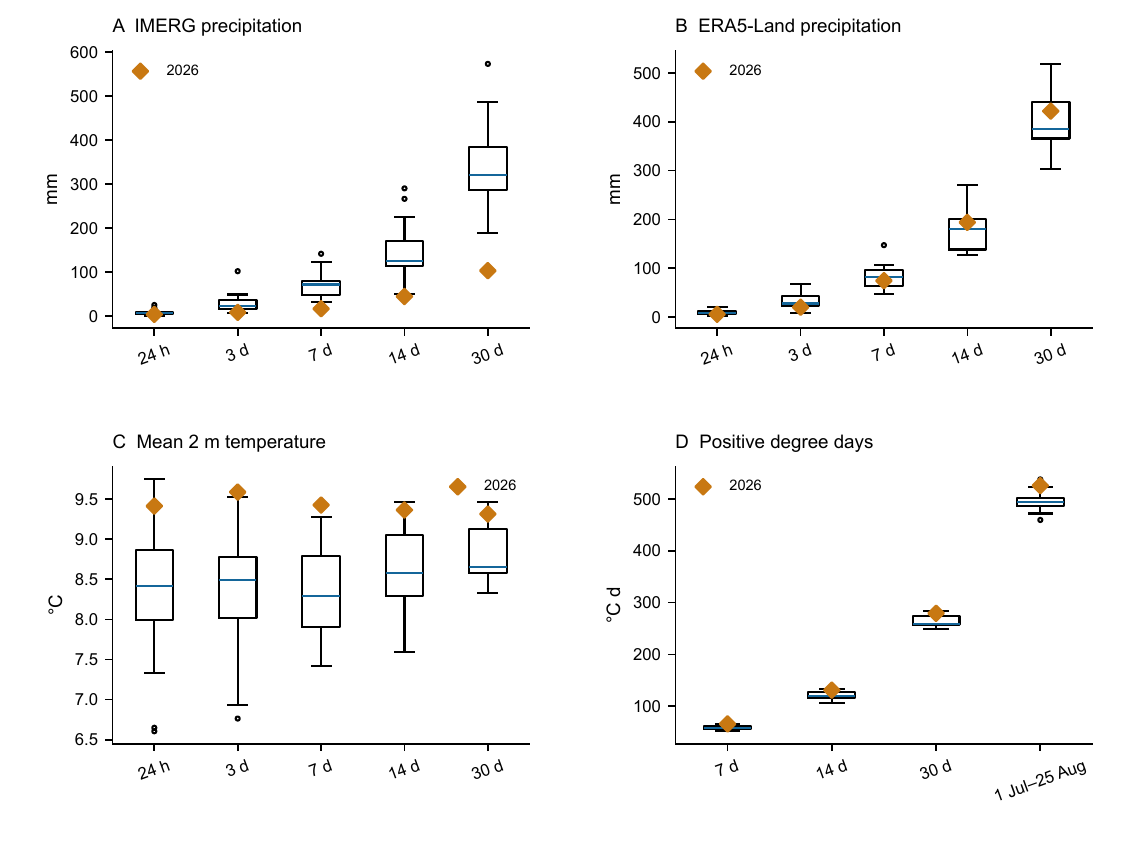}
\caption{\textbf{Matched-year source-nearest hydroclimate distributions.} (A,B) IMERG and ERA5-Land precipitation. (C) Hourly mean ERA5-Land 2~m temperature. (D) Positive degree days from complete UTC days. Boxes span the historical interquartile range, lines mark medians, whiskers extend to values within 1.5 interquartile ranges and dots mark outliers. Diamonds denote 2026, compared with the 25 matched years 2001--2025. Hourly windows end at their product cutoffs; positive degree days end on 25 August and are not identical temporal windows. All temperatures are unadjusted for elevation. Positive degree days are thermal indices, not measured melt. Overlapping windows are dependent. Data: NASA GPM IMERG and Copernicus Climate Change Service ERA5-Land.}
\label{fig:s3}
\end{figure}

\begin{figure}[!htbp]
\centering
\includegraphics[width=\textwidth,height=0.64\textheight,keepaspectratio]{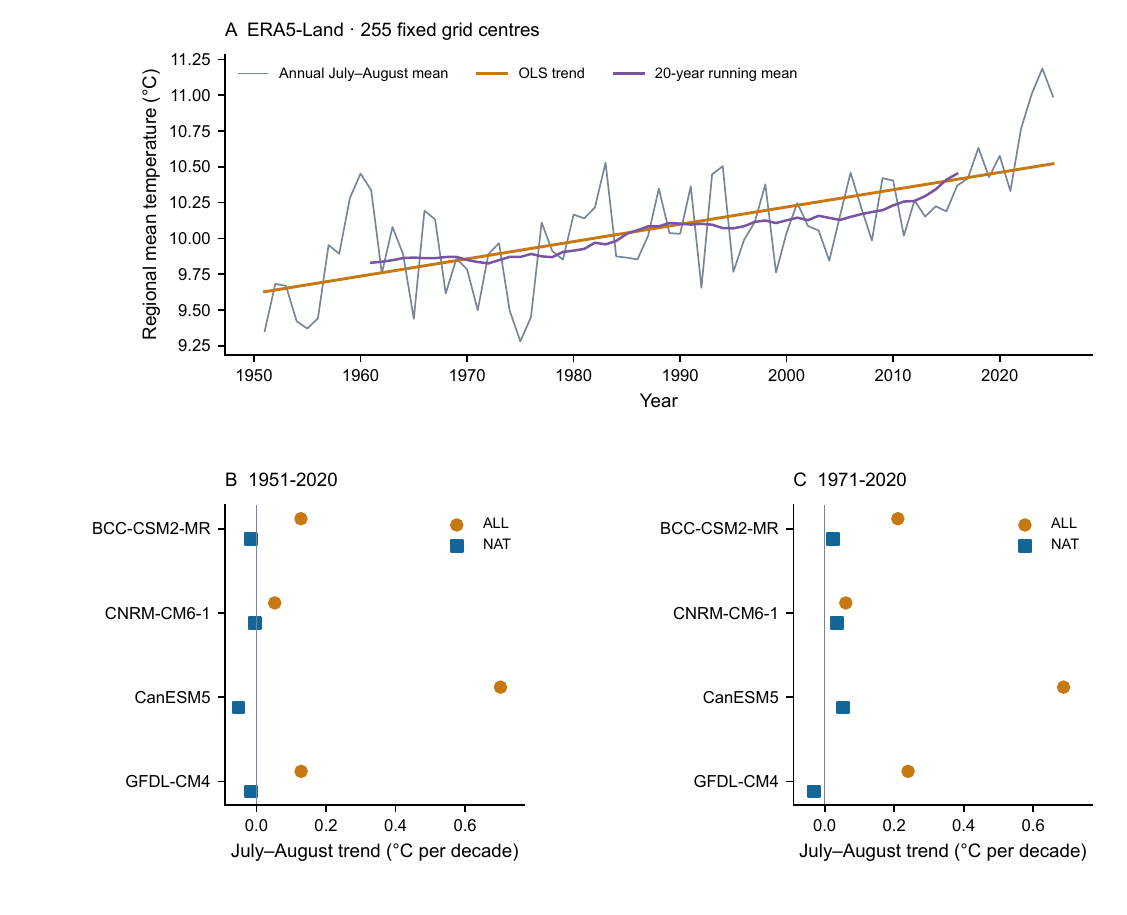}
\caption{\textbf{Regional warming and a limited model-forcing comparison.} (A) Annual July--August ERA5-Land regional means, a centred 20-year running mean and an ordinary least-squares trend over 1951--2025. This is a seasonal series, not annual-mean temperature. The fixed support comprises 255 centres over 27.6--29.2$^{\circ}$ N, 84.8--86.2$^{\circ}$ E. (B,C) Paired July--August trends from four models under all forcing (ALL: historical followed by SSP2-4.5 after 2014) and historical natural forcing (NAT). One member per model and four complete triplets were selected by archive availability. The model selection box is 27--30$^{\circ}$ N, 84--87$^{\circ}$ E and differs from the ERA5-Land region; CanESM5 contributes only one cell. Table~\ref{tab:s8} gives observational intervals, model values and grid supports. Model spread is not a probability distribution for natural variability. The comparison is descriptive and does not attribute the rupture to anthropogenic warming. Data: Copernicus Climate Change Service ERA5-Land and the named CMIP6/DAMIP modelling groups.}
\label{fig:s4}
\end{figure}

\FloatBarrier
\subsection*{Table S1. Interpreted source surfaces and terrain}\setcounter{table}{0}\refstepcounter{table}\label{tab:s1}
\begingroup\small
\setlength{\tabcolsep}{4pt}
\renewcommand{\arraystretch}{1.15}
\begin{longtable}{lrrrrrrr}
\toprule
Envelope & Area (km$^2$) & Elev. range & Median elev. & Slope & Aspect & Relief & Rectangle \\
\midrule
\endhead
Conservative & 0.491 & 4543--5200 & 5045 & 35.7 & 280.5 & 804 & 798 \\
Preferred & 1.009 & 4355--5221 & 4991 & 34.8 & 280.8 & 840 & 1151 \\
Inclusive & 1.841 & 4261--5261 & 4977 & 32.1 & 285.2 & 849 & 1526 \\
\bottomrule
\end{longtable}
All linear terrain measures are metres; slope and circular mean aspect are degrees. Aspect is clockwise from north. Relief is a 101$\times$101-cell square-window range on the 10~m grid. Rectangle is the minimum rotated rectangle long side. Rounded areas are shown; exact areas are given in Supplemental Materials and Methods.
\endgroup
\subsection*{Table S2. Exact optical acquisition identities}\setcounter{table}{1}\refstepcounter{table}\label{tab:s2}
\begingroup\footnotesize
\setlength{\tabcolsep}{4pt}
\renewcommand{\arraystretch}{1.15}
\begin{longtable}{llll}
\toprule
Panel & Sentinel-2 Level-2A scene ID & Time UTC & Usable (\%) \\
\midrule
\endhead
Fig. 2A & \texttt{S2A\_45RUM\_20260824\_1\_L2A} & 05:11:03.721 & -- \\
1 & \texttt{S2C\_45RUM\_20251006\_0\_L2A} & 05:11:03.655 & 85.9 \\
2 & \texttt{S2C\_45RUM\_20260213\_0\_L2A} & 05:10:50.634 & 100.0 \\
3 & \texttt{S2B\_45RUM\_20260516\_0\_L2A} & 05:00:49.693 & 100.0 \\
4 & \texttt{S2A\_45RUM\_20260702\_0\_L2A} & 05:01:09.386 & 77.7 \\
5 & \texttt{S2C\_45RUM\_20260812\_0\_L2A} & 05:10:48.070 & 42.3 \\
6 & \texttt{S2B\_45RUM\_20260824\_0\_L2A} & 05:00:49.736 & 62.2 \\
\bottomrule
\end{longtable}
The YYYYMMDD segment gives the acquisition date. Figure~\ref{fig:s2} rows 1--6 have complete source coverage. Usable fractions use the 20~m selection grid and covered preferred-source denominator; they do not apply to Figure~\ref{fig:2}A. Figure~\ref{fig:2}B uses Landsat-9 on 26 August 2026 at 04:47:48 UTC; Figure~\ref{fig:2}C uses Sentinel-2 on 27 August at approximately 05:10 UTC. Radar scene identifiers are S1D\_IW\_GRDH\_1SDV\_20260816T122141\_20260816T122206\_004151\_007980\_rtc and S1D\_IW\_GRDH\_1SDV\_20260828T122141\_20260828T122206\_004326\_007FA4\_rtc. The selected sequence ends at 05:00:49.736 UTC on 24 August; 45.85~h uses that time, not Figure~\ref{fig:2}A.
\endgroup

\subsection*{Table S3. Fixed-support precipitation totals and matched ranks}\setcounter{table}{2}\refstepcounter{table}\label{tab:s3}
\begingroup\footnotesize
\setlength{\tabcolsep}{4pt}
\renewcommand{\arraystretch}{1.15}
\begin{longtable}{lllrrrrr}
\toprule
Product & Support & Cells & 24 h & 3 d & 7 d & 14 d & 30 d \\
\midrule
\endhead
IMERG & Source & 1 & 4.0 (20) & 8.3 (4) & 16.8 (0) & 44.6 (0) & 103.4 (0) \\
IMERG & North & 1 & 3.6 (44) & 7.2 (16) & 18.3 (8) & 59.6 (12) & 164.4 (4) \\
IMERG & South & 1 & 6.9 (60) & 18.0 (16) & 37.4 (0) & 71.0 (0) & 137.7 (0) \\
IMERG & East & 1 & 2.3 (8) & 6.4 (4) & 13.4 (0) & 36.5 (0) & 92.7 (0) \\
IMERG & West & 1 & 4.7 (48) & 11.6 (8) & 26.6 (4) & 54.5 (0) & 131.3 (0) \\
IMERG & 3x3 mean & 9 & 4.3 (40) & 10.7 (8) & 24.9 (4) & 59.0 (4) & 140.0 (0) \\
ERA5-Land & Source & 1 & 5.7 (16) & 20.2 (16) & 74.5 (44) & 194.0 (64) & 422.0 (68) \\
ERA5-Land & North & 1 & 3.9 (16) & 15.7 (16) & 58.9 (40) & 153.6 (72) & 325.5 (68) \\
ERA5-Land & South & 1 & 8.4 (28) & 23.5 (12) & 85.0 (36) & 212.9 (48) & 480.8 (52) \\
ERA5-Land & East & 1 & 5.2 (16) & 19.5 (20) & 73.3 (56) & 181.8 (68) & 382.9 (68) \\
ERA5-Land & West & 1 & 6.4 (20) & 21.2 (12) & 74.5 (32) & 207.1 (60) & 470.0 (68) \\
ERA5-Land & 3x3 mean & 9 & 6.0 (20) & 19.9 (12) & 72.6 (40) & 187.8 (56) & 414.3 (64) \\
\bottomrule
\end{longtable}
Each entry is mm (empirical percentile) for 2001--2025, $n=25$. A rank of 0 means below all 25 matched years. Single-cell centres are defined by the source-nearest cell and its north, south, east and west neighbours in the following blocks. The ERA5-Land 3$\times$3 block has centres at 28.2, 28.3, 28.4$^{\circ}$ N and 85.4, 85.5, 85.6$^{\circ}$ E. The IMERG block has centres at 28.15, 28.25, 28.35$^{\circ}$ N and 85.45, 85.55, 85.65$^{\circ}$ E. Block means are arithmetic and their ranks use their own historical distributions. All retained event and historical windows are complete. IMERG Late historical/event versions are V07B/V07C and end at 02:30 UTC; ERA5-Land ends at 02:00 UTC and event data include preliminary ERA5-Land-T, expver 0005. All 60 ranks are below the declared 90th-percentile threshold. Version effects were not quantified.
\endgroup
\subsection*{Table S4. Seven-day temperature sensitivity}\setcounter{table}{3}\refstepcounter{table}\label{tab:s4}
\begingroup\small
\setlength{\tabcolsep}{4pt}
\renewcommand{\arraystretch}{1.15}
\begin{longtable}{lllrrrrl}
\toprule
Support & Cells & Centre N, E & Mean & Hist. median & Anomaly & Rank & Above all \\
\midrule
\endhead
Source & 1 & 28.30, 85.50 & 9.426 & 8.292 & +1.134 & 100 & Yes \\
North & 1 & 28.40, 85.50 & 8.429 & 7.414 & +1.016 & 96 & No \\
South & 1 & 28.20, 85.50 & 7.760 & 6.255 & +1.505 & 100 & Yes \\
East & 1 & 28.30, 85.60 & 5.171 & 3.163 & +2.009 & 100 & Yes \\
West & 1 & 28.30, 85.40 & 13.393 & 12.731 & +0.661 & 88 & No \\
3x3 mean & 9 & 28.30, 85.50 & 8.594 & 7.350 & +1.245 & 100 & Yes \\
\bottomrule
\end{longtable}
Mean, historical median and anomaly are $^{\circ}$C; rank is empirical percentile. ERA5-Land supports and block centres are as defined in Table~\ref{tab:s3}. Each baseline comprises 25 matched 7-day means over 2001--2025. Event and historical coverage is complete; the event uses preliminary ERA5-Land-T and ends 26 August at 02:00 UTC. All anomalies are positive, but only four supports exceed all historical means; the west is below the 90th percentile. There is no lapse-rate correction or local melt/bed-temperature inference.
\endgroup

\subsection*{Table S5. Fixed-zone area and historical built-up change}\setcounter{table}{4}\refstepcounter{table}\label{tab:s5}
\begingroup\small
\setlength{\tabcolsep}{4pt}
\renewcommand{\arraystretch}{1.15}
\begin{longtable}{lrrrrrrr}
\toprule
Zone & Zone km$^2$ & 1975 & 2020 & Net change & Rel. \% & 1975 \% & 2020 \% \\
\midrule
\endhead
A & 37.354195 & 0.035947 & 0.441243 & 0.405296 & 1127.5 & 0.096 & 1.181 \\
B & 1.258754 & 0.005408 & 0.036293 & 0.030885 & 571.1 & 0.430 & 2.883 \\
C & 1.110779 & 0.002575 & 0.024615 & 0.022040 & 855.9 & 0.232 & 2.216 \\
D & 22.438685 & 0.000728 & 0.009516 & 0.008788 & 1207.1 & 0.003 & 0.042 \\
\bottomrule
\end{longtable}
A: UNOSAT FloodExtent\_20260826\_Nepal union

B: Copernicus EMSR927 AOI02 observed-event fixed polygon

C: Copernicus EMSR927 AOI01 observed-event fixed polygon

D: 500 m buffer around representative source-to-port centreline\par Built-up areas and net changes are km$^2$. The final two columns are percent of the full fixed zone area. Relative change is $100(B_{2020}-B_{1975})/B_{1975}$ and is sensitive to small historical baselines; all baselines are nonzero. It is not a loss or risk multiplier. Areas use the exact raw-layer unions and EPSG:32645. Zones overlap and EMS polygons are not whole municipalities.
\endgroup
\clearpage
\subsection*{Table S6. All GHSL epochs and boundary sampling}\setcounter{table}{5}\refstepcounter{table}\label{tab:s6}
\begingroup\small
\setlength{\tabcolsep}{4pt}
\renewcommand{\arraystretch}{1.15}
\begin{longtable}{lrrrrr}
\toprule
Zone & Epoch & Preferred m$^2$ & Sampling band m$^2$ & Fraction \% & Fraction band \% \\
\midrule
\endhead
A & 1975 & 35947 & 16793--53719 & 0.096 & 0.045--0.144 \\
A & 1990 & 133584 & 69142--172542 & 0.358 & 0.185--0.462 \\
A & 2000 & 187522 & 95088--244794 & 0.502 & 0.255--0.655 \\
A & 2015 & 335715 & 176171--457665 & 0.899 & 0.472--1.225 \\
A & 2020 & 441243 & 230669--603316 & 1.181 & 0.618--1.615 \\
B & 1975 & 5408 & 3198--5917 & 0.430 & 0.254--0.470 \\
B & 1990 & 12201 & 5851--14001 & 0.969 & 0.465--1.112 \\
B & 2000 & 16671 & 8137--18779 & 1.324 & 0.646--1.492 \\
B & 2015 & 31510 & 17225--35812 & 2.503 & 1.368--2.845 \\
B & 2020 & 36293 & 20805--41294 & 2.883 & 1.653--3.281 \\
C & 1975 & 2575 & 867--3711 & 0.232 & 0.078--0.334 \\
C & 1990 & 14830 & 5555--19623 & 1.335 & 0.500--1.767 \\
C & 2000 & 19082 & 7235--25111 & 1.718 & 0.651--2.261 \\
C & 2015 & 24575 & 10330--31396 & 2.212 & 0.930--2.826 \\
C & 2020 & 24615 & 10349--31443 & 2.216 & 0.932--2.831 \\
D & 1975 & 728 & 625--883 & 0.003 & 0.003--0.004 \\
D & 1990 & 2048 & 1860--2270 & 0.009 & 0.008--0.010 \\
D & 2000 & 2674 & 2486--2896 & 0.012 & 0.011--0.013 \\
D & 2015 & 7669 & 7480--7898 & 0.034 & 0.033--0.035 \\
D & 2020 & 9516 & 9318--9751 & 0.042 & 0.042--0.043 \\
\bottomrule
\end{longtable}
Zone letters and full-area denominators are given in Table~\ref{tab:s5}. Preferred values sum native-cell square metres by cell-centre inclusion. Lower/upper alternatives select centres in a 65~m eroded core / all cells touched by the original polygon. Bands exclude classification and historical-model uncertainty. No pixel-area multiplier is applied. In every zone the 2020 lower estimate exceeds the 1975 upper estimate. Machine-readable data also include selected-cell counts and absolute/relative changes by epoch.
\endgroup

\subsection*{Table S7. Delivered damage accounting}\setcounter{table}{6}\refstepcounter{table}\label{tab:s7}
\begingroup\small
\setlength{\tabcolsep}{4pt}
\renewcommand{\arraystretch}{1.15}
\begin{longtable}{lllrr}
\toprule
Area & Feature / unit & Grade & Delivered total & Mappable points \\
\midrule
\endhead
Syapru Besi & Building points & Destroyed & 323 & 323 \\
Syapru Besi & Building points & Damaged & 32 & 32 \\
Syapru Besi & Building points & Possibly damaged & 78 & 78 \\
Syapru Besi & Roads (km) & Destroyed & 6.147756 & -- \\
Syapru Besi & Roads (km) & Possibly damaged & 1.485962 & -- \\
Syapru Besi & Roads (km) & No visible damage & 3.871929 & -- \\
Timure & Building points & Destroyed & 372 & 372 \\
Timure & Building points & Damaged & 33 & 33 \\
Timure & Building points & Possibly damaged & 26 & 26 \\
Timure & Roads (km) & Destroyed & 9.523160 & -- \\
Timure & Roads (km) & Possibly damaged & 0.175133 & -- \\
\bottomrule
\end{longtable}
All building records have map coordinates. Two Syapru Besi road records graded no visible damage lack geometry; the length for that grade uses available geometries only. Across both areas, 695 of 864 building records are graded destroyed. The destroyed-road total is 15.670916~km, measured in EPSG:32645. Six transport points and two facility polygons are additionally graded destroyed. These provider interpretations are not a complete asset census or field validation. Road lengths measure available delivered geometries. Source: Copernicus EMSR927 AOI01/AOI02.
\endgroup
\clearpage
\subsection*{Table S8. Regional temperature and model comparison}\setcounter{table}{7}\refstepcounter{table}\label{tab:s8}
\begingroup\small
\setlength{\tabcolsep}{4pt}
\renewcommand{\arraystretch}{1.15}
\begin{longtable}{llrr}
\toprule
ERA5-Land support & Period & OLS trend & 95\% OLS interval \\
\midrule
\endhead
Source & 1951--2025 & 0.0666 & 0.0318--0.1014 \\
Source & 1976--2025 & 0.0675 & 0.0137--0.1212 \\
Source & 2001--2025 & 0.3117 & 0.2135--0.4099 \\
Regional & 1951--2025 & 0.1208 & 0.0894--0.1521 \\
Regional & 1976--2025 & 0.1482 & 0.0954--0.2011 \\
Regional & 2001--2025 & 0.3522 & 0.2291--0.4752 \\
\bottomrule
\end{longtable}
Trends and intervals are $^{\circ}$C per decade for July--August. OLS intervals assume independent residuals, exclude serial-dependence correction and do not include full product uncertainty. The short-term source cell, ERA5-Land regional support and larger Berkeley/CMIP selection box are different. The continuation below reports comparison products and models.
\endgroup

\subsection*{Table S8 continued}
\begingroup\small
\setlength{\tabcolsep}{4pt}
\renewcommand{\arraystretch}{1.15}
\begin{longtable}{lrr}
\toprule
Period & Berkeley Earth trend (95\% CI) & ERA5-Land trend (95\% CI) \\
\midrule
\endhead
1951-2024 & 0.1531 (0.1128--0.1933) & 0.1156 (0.0840--0.1472) \\
1976-2024 & 0.2597 (0.1951--0.3243) & 0.1378 (0.0842--0.1914) \\
2001-2024 & 0.3691 (0.1824--0.5557) & 0.3313 (0.2000--0.4627) \\
\bottomrule
\end{longtable}
Berkeley Earth uses nine 1$^{\circ}$ cells and cosine-latitude weighting. ERA5-Land uses 255 0.1$^{\circ}$ centres and arithmetic weighting. Differing support sizes preclude interpreting numerical equality as a validation target.
\endgroup
\begingroup\small\begin{center}\begin{tabular}{llrrrr}\toprule Model & Member / grid & Cells & Period & ALL & NAT \\ \midrule
BCC-CSM2-MR & r1i1p1f1/gn & 9 & 1951-2020 & 0.1277 & -0.0166 \\
CNRM-CM6-1 & r1i1p1f2/gr & 4 & 1951-2020 & 0.0523 & -0.0050 \\
CanESM5 & r1i1p1f1/gn & 1 & 1951-2020 & 0.7016 & -0.0519 \\
GFDL-CM4 & r1i1p1f1/gr1 & 9 & 1951-2020 & 0.1282 & -0.0153 \\
BCC-CSM2-MR & r1i1p1f1/gn & 9 & 1971-2020 & 0.2104 & 0.0240 \\
CNRM-CM6-1 & r1i1p1f2/gr & 4 & 1971-2020 & 0.0608 & 0.0356 \\
CanESM5 & r1i1p1f1/gn & 1 & 1971-2020 & 0.6870 & 0.0534 \\
GFDL-CM4 & r1i1p1f1/gr1 & 9 & 1971-2020 & 0.2399 & -0.0307 \\
\bottomrule\end{tabular}\end{center}
ALL continues historical forcing with SSP2-4.5 after 2014; NAT is historical-natural. One member per model; four complete triplets selected by archive availability. Model grid centres and archive identifiers are in the processed data. Observed ERA5-Land trends for the two comparison periods are 0.0966 and 0.1252$^{\circ}$C per decade. None of the ALL point trends falls within the observed 95\% OLS interval for 1971--2020. The range is descriptive, not a confidence interval or event-attribution test.\endgroup
\clearpage
\subsection*{Table S9. Fixed disturbed-footprint transects}\setcounter{table}{8}\refstepcounter{table}\label{tab:s9}
\begingroup\small
\setlength{\tabcolsep}{4pt}
\renewcommand{\arraystretch}{1.15}
\begin{longtable}{lrllcc}
\toprule
Transect & Component (m) & Status & Crossings & Start inside & End inside \\
\midrule
\endhead
T01 & $\geq$1097.651 & censored & 1 & False & True \\
T02 & $\geq$1069.250 & censored & 1 & False & True \\
T03 & $\geq$1165.155 & censored & 1 & True & False \\
T04 & $\geq$1298.044 & censored & 1 & True & False \\
T05 & 970.330 & complete & 2 & False & False \\
T06 & 677.312 & complete & 2 & False & False \\
T07 & 388.902 & complete & 2 & False & False \\
T08 & 219.673 & complete & 2 & False & False \\
T09 & 286.878 & complete & 2 & False & False \\
T10 & 281.806 & complete & 2 & False & False \\
T11 & 145.151 & complete & 2 & False & False \\
T12 & 195.230 & complete & 2 & False & False \\
\bottomrule
\end{longtable}
T01--T04 are lower bounds because one endpoint remains inside the selected footprint component. Every lower bound exceeds the two complete values at ranks 6 and 7: T07 = 388.902453~m and T06 = 677.312079~m. Their average gives the identifiable median, 533.107266~m. No finite overall maximum or complete-width range is inferred. Directions and component-selection rules are fixed; these are not water or hydraulic widths.
\endgroup

\begingroup
\small
\setlength{\parskip}{0pt}
\bibliographystyle{unsrt}
\bibliography{references}
\endgroup
\end{document}